\documentclass[twocolumn,epsfig,address]{revtex4}
\usepackage{graphicx}
\usepackage{amssymb}
\usepackage{bm}
\usepackage{mathptm,amsmath} 

\renewcommand{\vec}{\mathbf}

\begin{document}

\title{Poisson-Boltzmann for oppositely charged bodies: an explicit derivation}

\author{Fabien Paillusson, Maria Barbi, Jean-Marc Victor}
\affiliation{Laboratoire de Physique Th\'{e}orique de la Mati\`{e}re 
Condens\'{e}e, Universit{\'{e}} Pierre et
Marie Curie, case courrier 121, 4 Place Jussieu - 75252 Paris 
cedex 05, France }
\date{\today}

\begin{abstract}
{
The interaction between charged bodies in an ionic solution is a
general problem in colloid physics and becomes a central topic in the
study of biological systems where the electrostatic interaction
between proteins, nucleic acids, membranes is involved. This problem
is often described starting from the simple one-dimensional model of
two parallel charged plates.  Several different approaches to this
problem exist, focusing on different features.  In many cases, an
intuitive expression of the pressure exerted on the plates is
proposed, which includes an electrostatic plus an osmotic
contribution. We present an explicit and self-consistent derivation of
this formula for the general case of any charge densities on the
plates and any salt solution, obtained in the framework of the
Poisson-Boltzmann theory. We also show that, depending on external constraints,
the correct thermodynamic potential can differ from the usual PB free energy. 
The resulting expression predicts, for asymmetric, oppositely charged
plates, the existence of a non trivial equilibrium position with the
plates separated by a finite distance. It is therefore crucial, in
order to study the kinetic stability of the corresponding energy
minimum, to obtain its explicit dependence on the plates charge
densities and on the ion concentration. An analytic expression for the
position and value of the corresponding energy minimum has been
derived in 1975 by Ohshima [{\bf Ohshima H., Colloid and Polymer
Sci. 253, 150-157 (1975)}] but, surprisingly, this important result
seems to be overlooked today. We retrieve the expressions obtained by
Ohshima in a simpler formalism, more familiar to the physics
community, and give a physical interpretation of the observed
behavior.  
}%
\end{abstract}

\maketitle

\section{Introduction}

Poisson-Boltzmann theory is a statistical mean field theory that
characterizes coarse-grained quantities such as the average particle
distribution function and the electrostatic potential together with
thermodynamic variables in systems composed of many charged and point
like particles at thermal equilibrium. Despite the technical advances
in the dilute and strong coupling regime \cite{Lau1,Orl2,Orl3}, the
statistical modeling of real solutions -- often in an {\em intermediate}
regime -- is still an open problem \cite {Orl1}. The PB approximation
remains a good reference theory for describing the essential features
of electrolyte solutions at thermal equilibrium.  It allows to model
plasmas in the equilibrium regime, colloidal suspensions through the
famous Cell Model \cite{Deserno}, or polyelectrolytes in solution.
Moreover, the increasing interest for the biological mechanisms at the
sub-cellular scale leads the community to deal with the electrostatic
interaction of biological objects in solution, as for the case of
protein-protein interaction \cite{protprot}, protein-DNA interaction
\cite{von07}, DNA-membrane interaction \cite{Joanny}, etc.

In the case where one is interested in the effective interaction
between two charged bodies surrounded by mobile charges, it is
frequently useful, given the difficulty of the equations that have to
be solved, to rely on a one dimensional problem to capture the physics
of the system \cite{Parsegian}. This essentially amounts to focus on
the interaction between two parallel charged plates in solution.
Besides, approximated methods have been developed in the past century
to correct the 1D problem as to take into account the geometric
effects in the interaction of two mesoscopic bodies, thus increasing
all the more the interest of one dimensional models \cite{verwey,
heli1, heli2, Tamashiro, Zypman}.

In general, the main quantities to be derived in the one dimensional
case are {\it (i)} an expression for the free energy of the system in
the framework of the Poisson-Boltzmann approximation, {\it (ii)} a
differential equation for the mean electrostatic potential and, in
order to evaluate the actual interaction between the two plates, {\it
(iii)} an explicit expression for the pressure exerted on each
surface.

Various derivations of the Poisson-Boltzmann approximation actually
exist. A good review of many ways to obtain the Poisson-Boltzmann
equation has been presented by Lau \cite{Lau} including a saddle point
approximation in a path integral formulation (see also
\cite{Orl1}). Less straightforward derivations are also available via
the Density Functional Theory (DFT) \cite{hansen} or exact equations
hierarchy \cite{Carnie}. Finally a less formal procedure has been
proposed by Deserno {\em et al.}, in the field of colloid physics, to
obtain mean field quantities for charged systems \cite{Deserno}.

Most presentations, despite their different approaches, lead to a same
formula for the pressure, which amounts to the sum of a purely
electrostatic plus a purely osmotic contribution. One merit of the
Poisson-Boltzmann approximation is indeed that this formula exactly
matches the boundary-density theorem at the Wigner-Seitz cell boundary
\cite{Marcus} as well as the contact value theorem on the charged
plates \cite{Wen82}. The first question addressed in this paper is
thus whether or not this intuitive expression for the inter-plate
pressure can be directly and exactly derived from the
Poisson-Boltzmann free energy, without need for additional arguments
and for any boundary conditions.  After having introduced the system
and its Poisson Boltzmann free energy in Section~\ref{freeenergy}, we
derive in Section~\ref{pbpressure} the expected expression for the
pressure and show that a particular caution should be taken in the
choice of the right statistical ensemble when different ``external''
constraints are imposed to the plates, as e.g. at constant potential
or at constant charge conditions.

The pressure formula predicts the presence of a non trivial
equilibrium distance for plates of opposite and asymmetric charge
densities.  This has been shown in the pioneering work of Parsegian
and Gingell \cite{Parsegian} who used the linear Debye-H\"ukel theory
in the case of high salt concentrations, and more recently, by Lau and
Pincus \cite{Lau99} in the framework of the nonlinear Poisson-Boltzmann
equation restricted to the case of no added salt.

{ The consequences of such an equilibrium on the effective behaviour
of charged bodies in solution can only be assessed by a study of the
corresponding energy profile, i.e. a comparison of the energy well
depth to the thermal energy. If the energy gain at the minimum is
small with respect to $k_B T$, the two charged bodies will not
stabilize in the bound complex and will behave as in the absence of
electrostatic interaction. Quite surprisingly, this aspect of the
problem is rarely addressed in the contemporary literature. Some
authors \cite{Ben-Yaakov} discuss in details how the equilibrium
distance (the limit between attraction and repulsion) depends on the
plate charges and on the salt conditions, but do not address the
question of the depth of the free energy well. Nevertheless, very nice
analytic expressions for both the position and the energy values at
the equilibrium position have been obtained in 1975 by Ohshima
\cite{Ohs75}.  The paper by Ohshima deals with the more complex case
of two parallel plates of given thickness and dielectric constant,
thus leading to a rather complex notation. Nonetheless, the important
results of Ref.~\cite{Ohs75} are worth being reproduced today at least
in the more usual case of two charged surfaces, in that they represent
an exact and synthetic description of their interaction whatever their
charges and the ionic strength of the solution.

In order to illustrate the system behavior in the simple but crucial
case of monovalent solutions, in Section~\ref{Ohshima} we first solve
explicitly the Poisson-Boltzmann problem and obtain the pressure and
energy profiles.  Then, we focus on the origin of the energy minimum
and derive an expression for its position and depth in the framework
of the Poisson-Boltzmann theory.  We check the agreement between the
analytic expression and the behaviour obtained by direct numerical
integration of the Poisson-Boltzmann equation. Finally, we discuss the
physical origin of the results by investigating the role of the
different parameters, as the plate charges and the salt ions and
counter-ions.  }%

\section{The Poisson-Boltzmann free energy of the two plates system}
\label{freeenergy}

{
We are interested in the thermodynamic properties of a system composed
of a fixed distribution of charges and of $N$ point-like mobile ions
in a solution at temperature $T$. The valence, mass, position and
momentum of the ion indexed by ``$i$'' are denoted by $z_{i}$, $m_i$,
$r_i$ and $p_i$, respectively. The Hamiltonian of the system can be
written as follows:
\begin{eqnarray}
H(\{\vec{r}\},\{\vec{p}\}) &=& H_{kin}+H_{pot} =\nonumber\\ 
= \sum_{i=1}^{N}\frac{p_{i}^{2}}{2m_{i}}
&+& \frac{1}{2} \sum_{i=1}^N z_i e \phi(\vec{r}_i) 
      +  \frac{1}{2}\int  e \sigma(\vec{r}) \phi(\vec{r}) d^3 r
\label{eqnH}
\end{eqnarray}
where $\sigma$ is the fixed volumic charge distribution in unit of the
elementary charge $e$ and $\epsilon=\epsilon_0 \epsilon_r$ is the
dielectric constant of the solvent. 
The function $\phi(\vec{r})$ is 
the electrostatic potential,
\begin{eqnarray}
\phi(\vec{r}) & \equiv & \sum_{j=1}^N \frac{z_j e}{4 \pi \epsilon |\vec{r}-\vec{r}_j|} 
+ \int \frac{\sigma(\vec{r}\:') e}{4\pi \epsilon |\vec{r}-\vec{r}\:'|}  d^3r'
\nonumber \\
&=& \sum_{\alpha=1}^p \int \frac{ e z_{\alpha} n_{\alpha}(\textbf{r}') }
{4 \pi \epsilon |\vec{r}-\vec{r'}|} d^3r'
+ \int \frac{\sigma(\vec{r}\:') e}{4\pi \epsilon |\vec{r}-\vec{r}\:'|}  d^3r' \,,
\label{eqn313}
\end{eqnarray}
where we introduced the ion density of the species $\alpha$,
$n_{\alpha}(\vec{r}) \equiv \sum_{i=1}^{N_{\alpha}}
\delta(\vec{r}-\vec{r}_i)$, and $\sum_{\alpha=1}^{p}N_{\alpha}=N$.

\begin{figure}[ht]
  \begin{center}
  \includegraphics[width=0.4\textwidth]{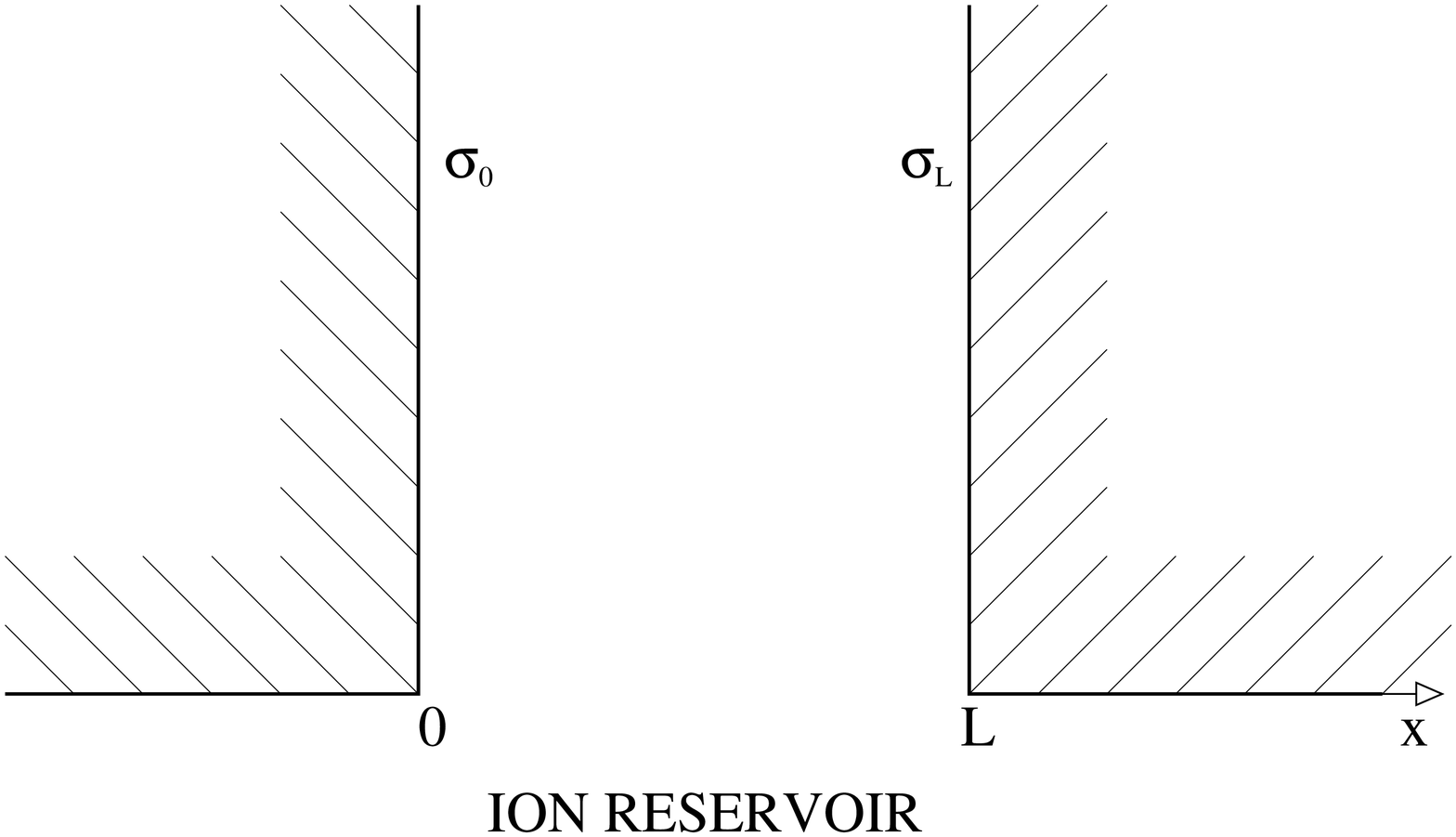}   
\caption{A schematic view of the system considered
throughout the paper. The two semi-infinite planes of charge density
$\sigma_0$ and $\sigma_L$, positioned respectively at $x=0$ and $x=L$,
are immersed in the ionic solution.  An ion reservoir freely exchanges
ions with the system.}
\label{fig:config} 
\end{center}
\end{figure}

We will then consider a system composed of two uniformly charged
plates separated by a distance $L$ and the electrolyte solution
between them (see Fig.~\ref{fig:config}).  The fixed charge
distribution is then
\begin{equation}
 \sigma(x)= \sigma_0 \delta(x) + \sigma_L \delta(x-L) 
 \label{eqnsigma}
\end{equation}
where $\sigma_0$ and $\sigma_L$ correspond to the surface charge
densities of the two plates positioned respectively at $x=0$ and
$x=L$.  The system is in contact with an infinite salt reservoir.
Only the $x$ coordinate is relevant due to the translation invariance
along the $z$ and $y$ directions.  In the following, we will focus on
the volume delimited by a given finite surface $A$ of the two facing
plates.

As usual, we can obtain the free energy of the system from the system
partition function $Z$, as $F\equiv -k_{B}T\ln Z$. The kinetic part
$Z_{kin}$ can be easily calculated \cite{Huang} and reads $Z_{kin} =
\prod_{\alpha=1}^{p} ( \Lambda_{\alpha}^{-3N_{\alpha}}/ {N_{\alpha}!})$,
where $\Lambda_{\alpha}\equiv{h}\, {({2\pi m_{\alpha}k_{B}T})}^{-1/2}$
is the de Broglie thermal wavelength.  The potential part of the
partition function is not that simple to compute, because the
electrostatic part of the Hamiltonian is a function of the position of
all ions and cannot reduce to a product of uncorrelated functions. The
simplest method to solve the problem is to rely on a mean field
approximation.  The Gibbs-Bogoliubov inequality allows one to find an
upper bound for the Helmoltz free energy from an average of the
Hamiltonian with a trial distribution $P_0(\vec{x})$ plus a Shannon
type entropy built from the same distribution $P_0(\vec{x})$. For a
given surface $A$, one gets therefore the following expression for the
free energy functional per unit surface -- that is the Poisson
Boltzmann free energy functional:
\begin{eqnarray}
 \frac{F_{PB}[\{n^0\}]}{A}&=& \lim_{h \to 0} \; 
\frac{1}{2}\int_{-h}^{L+h} \rho^0(x) \phi^0(x) dx  \\
&+& k_B T \sum_{\alpha=1}^p \int_{-h}^{L+h} 
\left\lbrace  n_{\alpha}^0(x)
\left( \ln\left( \Lambda_{\alpha}^{3} n_{\alpha}^0(x) \right)-1 \right)
\right\rbrace dx \,,
\nonumber
\label{eqnF-PB}
\end{eqnarray}
where we have introduced $n^0_{\alpha}(x) \equiv N_{\alpha}
P_0(x)$ following the normalization relation
$N_{\alpha}\equiv A \int n_{\alpha}^{0}(x) dx$, and the global
charge density $\rho^0$ defined by
\begin{equation}
 \rho^0(x)=\sum_{\alpha=1}^p z_{\alpha} e n^0_{\alpha}(x) + e \sigma(x) \,. 
\label{eqnrho}
\end{equation} 
 
We recognize, in the first term of this functional, the electrostatic
part of the energy of the system, while the second term corresponds to
the entropic contribution of an ideal gas of ions.

We should therefore minimize the functional ${F_{PB}[\{n^0\}]}/{A}$
with respect to the relevant functions $n_{\alpha}^0$. In order to
take into account properly the boundaries at $x=0$ and $x=L$ we
introduced a parameter $h$ for the calculations and then take the
limit for $h\to 0$.  The minimization should be performed under the
condition of conservation of the whole number of ions of type $\alpha$
in the system: this leads to define a {\em generalized} free energy
functional per unit area,
\begin{equation}
f_{PB} =  \frac{F_{PB}[\{n^0\}]}{A} 
- \lim_{h \to 0} \sum_{\alpha=1}^p \mu_{\alpha} \int_{-h}^{L+h} n^0_{\alpha}(x) dx 
\label{eqncalF}
\end{equation}
where $\mu_{\alpha}$ is a Lagrange multiplier that corresponds to the
electrochemical potential
\footnote{\label{fnote}The parameter $\mu_{\alpha}$ can
be obtained easily through the electrochemical potential of an ion of
type $\alpha$ in the salt reservoir. Indeed, in the electrolyte
solution of the infinite salt reservoir, coarse grained variables such
as the ion distribution $n^0_{\alpha}$ can be calculated by modeling
the system as a mixture of ideal gases. We have therefore
$\mu^{\Sigma_2}_{\alpha}=k_BT
\ln(c_{\alpha} \Lambda^3_{\alpha})$ (where $c_{\alpha}=N_{\alpha}/V$). 
As $\mu_{\alpha}^{\Sigma_2}= \mu_{\alpha}^{\Sigma_1}$ at equilibrium
we have $\mu_{\alpha}=k_BT \ln(c_{\alpha} \Lambda^3_{\alpha})$.
}  
of the ion type $\alpha$ in the reservoir. 

The minimization leads to the following relation between the {\em mean
field} ion distributions $\overline{n}_{\alpha}$ {\em minimizing}
$f_{PB}$ and the corresponding mean field potential
$\overline{\phi}(x)$:
\begin{equation}
 \overline{n}_{\alpha}(x)=\Lambda_{\alpha}^{-3} 
e^{\beta \mu_{\alpha}} e^{-\beta e z_{\alpha}\overline{\phi}(x)} \,.
\label{eqn16}
\end{equation}
The reader will recognize in this result an explicit expression of the
Boltzmann law, here rigorously re-obtained in the framework of the
mean field approach.

Together with Eq.~\eqref{eqn313}, giving the electric field as a
function of the charge distribution in the system, the previous
Equation~\eqref{eqn16} constitute the solution of the
problem. Eq.~\eqref{eqn16} allows to obtain a simpler expression for
the potential $\overline{\phi}(x)$ in terms of the free and fixed
charge distributions in the system.  Recalling that the electric
potential and the charge density are linked by the Poisson equation,
i.e.
\begin{equation}
 \Delta \phi^0 (\vec r) = - \frac{\rho^0(\vec r)}{\epsilon} \,
\label{eqn17}
\end{equation}
and combining with Eq.~\eqref{eqn16}, we obtain indeed an ordinary
differential equation for the adimensional mean field potential
$\overline{\psi}(x)~=~\beta e
\overline{\phi}(x)$. 
The resulting {\em Poisson-Boltzmann} (PB) equation  reads in
our one-dimensional case:
\begin{equation}
    \frac{d^2 \overline{\psi }(x)}{dx^2} = - 4 \pi \ell_B
    \sum_{\alpha=1}^p z_{\alpha} \overline{n}_{\alpha}(x),
\;\;\; x\in\lim_{h \to 0} [+h,L-h]
\label{eqn20} 
\end{equation}
with the boundary conditions
\begin{eqnarray}
\lim_{h \to 0}  \left.    \frac{d\overline{\psi}}{dx}\right|_{+h}=
-  4 \pi \ell_B  \sigma_0 \,, \nonumber \\
\lim_{h \to 0} \left.     \frac{d\overline{\psi}}{dx}\right|_{L-h}=
      4 \pi \ell_B  \sigma_L \,.
\label{eqnBC}
\end{eqnarray}
where $\ell_B\equiv {e^2}/{4\pi \epsilon k_B T}$ denotes the Bjerrum
length and where we used electroneutrality of the considered system
toget the boundary condition.

Before going on, we should note that it is possible to get a constant
of motion $C$ by multiplying ($\ref{eqn20}$) by
$\frac{d\overline{\psi}}{dx}$ and then integrating in the $[+h,L-h]$
range. One gets
\begin{eqnarray}
&& \frac{1}{2} \left( \frac{d\overline{\psi}}{dx} \right)^2
-4\pi \ell_B \sum_{\alpha=1}^p \overline{n}_{\alpha}(x)= C \,. 
\label{eqnE}
\end{eqnarray}
This result will have a crucial role in the definition of the pressure
between the two plates as we will see in the next section.

In order to compute a general expression for the pressure from a
thermodynamic definition, we have now to evaluate the PB functional
free energy at ${n}_{\alpha}^0(x) = \overline{n}_{\alpha}(x)$. The
result, written in an equivalent but more practical form, will be
identified with the system free energy.  Indeed, at the $h \to 0$
limit, the integrals of any non diverging function in the two external
regions vanish. Thus, after an integration by parts for the
electrostatic contribution, we obtain for the PB free energy expressed
in terms of the adimensional field $\overline{\psi}$:
\begin{eqnarray}
&&\beta f_{PB}[\{\overline{n}\}, \sigma]
=\lim_{h \to 0} \left\lbrace \frac{1}{8 \pi \ell_B}
 \right. 
 \int_{+h}^{L-h} 
\left( \frac{d\overline{\psi}}{dx}\right)^2 dx
 \label{eqncalF2}  \\
&&\hspace{6mm} + \sum_{\alpha=1}^p \int_{+h}^{L-h} \overline{n}_{\alpha}(x)  
\left.
 \left[ \ln\left( \Lambda^3_{\alpha} \overline{n}_{\alpha} \right) -1 
- \beta \mu_{\alpha} \right] dx \right\rbrace \,, \nonumber
\end{eqnarray}
where we used Eqs.~\eqref{eqnBC}.

\section{Determination of the pressure}
\label{pbpressure}

We can now address the problem of finding an explicit expression for
the pressure $\Pi$ on the plates, according to the Poisson-Boltzmann
theory.  In the case of two {\em constant} charged densities on the
plates, i.e. two plates whose charge densities are fixed once for
ever, the plates self energy is independent of $L$ and the usual
derivation of the pressure from the free energy can therefore be used:
\begin{equation}
 \Pi \equiv - A\, \left.
\frac{ \partial { f}_{PB}[\{\overline{n}_{\alpha}\}]}{\partial V}\right|_{\sigma} 
 = -\left.\frac{ \partial {f}_{PB}[\{\overline{n}_{\alpha}\}]}{\partial L}\right|_{\sigma}
 \label{eqn26}
\end{equation}

To facilitate the calculation, let introduce the adimensional variable
$\xi={x}/{L}$, \cite{Lau} and define a rescaled electric field ${\cal
E}$ for the adimensional potential, $ {\cal E} \equiv -\frac{\partial
\overline{\psi}}{\partial x}$. We then get directly from
($\ref{eqn20}$)
\begin{equation}
 \sum_{\alpha=1}^p z_{\alpha} \overline{n}_{\alpha}
 =\frac{1}{4\pi \ell_B} \frac{1}{L} \frac{\partial {\cal E}}{\partial \xi} \,. 
 \label{eqn29}
\end{equation}
The entropic part of \eqref{eqncalF2} can be written as
\begin{eqnarray}
s_{PB}[\{\overline{n}\}, \sigma]&=&
\lim_{h \to 0} \int_{+h}^{1-h} 
\Big( - \frac{\overline{\psi}}{4\pi \ell_B}\frac{1}{L} \frac{\partial {\cal E}}{\partial \xi} 
- \sum_{\alpha=1}^p \overline{n}_{\alpha} \Big) L \, d\xi 
\label{eqn30}
\end{eqnarray}
by using Eq.~\eqref{eqn16}. We can now take the derivative of
Eq.~\eqref{eqn30} with respect to $L$, and get
\begin{eqnarray}
\frac{\partial}{\partial L} s_{PB}[\{\overline{n}\}, \sigma]&=& \lim_{h \to 0} \int_{+h}^{1-h} 
\left[ -\frac{1}{4\pi \ell_B}\left( \frac{\partial \overline{\psi}}{\partial L}\frac{d {\cal E}}{d\xi}
+\overline{\psi} \frac{\partial}{\partial L}\frac{\partial {\cal E}}{\partial \xi}\right) \right. 
\nonumber \\
&& \left. \hspace{1.4cm}
-L \frac{\partial}{\partial L} \Big(\sum_{\alpha=1}^p \overline{n}_{\alpha}\Big)
-\Big( \sum_{\alpha=1}^p \overline{n}_{\alpha} \Big) \right]d\xi\nonumber \\
\label{eqnderiv}
\end{eqnarray}

Finally, using again Eq.~\eqref{eqn16} and recalling that
$\overline{n}_{\alpha}$ are functions of $x = \xi L$, the derivative
of the entropic term in the free energy, Eq.~\eqref{eqnderiv}, becomes
\begin{eqnarray}
 \lim_{h \to 0} \int_{+h}^{1-h}
\Big(-\frac{1}{4\pi \ell_B}\overline{\psi}\frac{ \partial }{\partial L} 
\frac{\partial {\cal E}}{\partial \xi}
-\sum_{\alpha=1}^p \overline{n}_{\alpha} \Big)\, d\xi \nonumber
\end{eqnarray}

In the same way, we calculate the partial derivative of the
electrostatic part of Eq.~\eqref{eqncalF2}.  Grouping the previous
results and using Eq.~\eqref{eqnE}, one gets for the pressure:
\begin{eqnarray}
\Pi= -\frac{k_B T}{4 \pi \ell_B} C + \lim_{h \to 0} \frac{k_B T}{4 \pi \ell_B}
\left[ \overline{\psi} \frac{\partial {\cal E}}{\partial L} \right]_{\xi=+h}^{\xi=1-h} 
\label{last0}
\end{eqnarray}
In the present case of constant charge densities on the plates, the
electric field at the boundaries is independent of $L$ and then the
second term of Eq. (\ref{last0}) vanishes. The final expression for
the pressure is therefore:
\begin{equation}
 \Pi=-\frac{k_B T}{4 \pi \ell_B} C =  -\frac{k_B T}{8 \pi \ell_B}
\left( \frac{d\overline{\psi}}{dx} \right)^2
+k_B T \sum_{\alpha=1}^p \overline{n}_{\alpha}(x) \label{last}
\end{equation}

The latter result is quite intuitive since it represents the sum of
the electrostatic stress and the osmotic pressure. It is indeed widely
used in the literature. However, in the general case of {\em non
constant} plate charge densities, the second term in Eq. \eqref{last0}
is a priori non zero. Such a term arises for instance when the
potential on the plates is kept constant. Feynman \cite{Feynman} already pointed out
this issue in it's famous course on electromagnetism for the case of
the pressure between the two parallel plates of a capacitor. If, for a
given distance $L$ between the plates and a given potential
difference, we try to evaluate the pressure by differentiating the
energy of the capacitor $CU^2/2$ with respect to $L$, then we get two
different results whether the differentiation is done at constant
charge or at constant potential. Of course the two derivations should
give the same result since they refer to the same state of the
system. The explanation of this apparent paradox is actually quite
simple: when differentiating at fixed potential we include implicitly
the energy supplied by the generator to keep the potential difference
constant while varying $L$. Since this work only modifies the self
energy of the plates and not their interaction, we have to subtract
this part from the result. We retrieve in this way the correct result
of the constant charge case.

In our case, one can easily realize that, at the boundaries, the
expression $\overline{\psi}\, \partial {\cal E}/\partial L$ is
equivalent to $\overline{\psi}\, \partial \sigma /\partial L$ and thus
corresponds to the energy used to bring charges to the plates when
varying the distance $L$. Then, the second term that arises in
Eq.(\ref{last0}) is exactly the analogous of the generator term in the
capacitor problem, and appears only because, using Gibbs terminology,
we don't work in the correct statistical ensemble. Indeed
Eq.(\ref{eqn26}) represents a very useful and systematic procedure to
get an expression for the pressure {\em provided} that the ensemble or
equivalently the effective potential is chosen carefully. For
instance, in the case of constant potential on the plates we have to
make a functional Legendre transform to get the relevant thermodynamic
potential:
\begin{equation}
g_{PB}[\{\overline{n}\},\{\overline{\psi}\}_b]=
f_{PB}[\{\overline{n}\},\sigma]-
\lim_{h \rightarrow 0} \int_{-h}^{L+h} \overline{\psi}(x) \sigma(x) dx \label{epi0}
\end{equation}
where $\{\overline{\psi}\}_b$ denotes the electrostatic potential at
the boundaries.  In this case Eq.~\eqref{eqn26} must be replaced by:
\begin{equation}
\Pi \equiv -\left.\frac{ \partial g_{PB}[\{\overline{n}_{\alpha}\}]}
{\partial L}\right|_{\{\overline{\psi}\}_b}.
\end{equation} 
Since the derivative of the additional term in $g_{PB}$ exactly
balances the second term of Eq.~\eqref{last0}, we finally retrieve the
general result \eqref{last}. For a different but equivalent discussion, see 
also Ref.\cite{Tri01}.

\section{Existence and characterization of the energy minimum for 
asymmetric charged plates}
\label{Ohshima}

\subsection{Zero pressure distance $L_{min}$}

It can be interesting to illustrate the implications of
Eq.~\eqref{last} for the case of two charged bodies in a 1:1
solution. The extension to a multivalent solution is straightforward
\footnote{Note however that the Poisson-Boltzmann approach is known to
be less accurate in the case of multivalent salt solutions. It has
been shown e.g. that a qualitatively different behavior can appear in
the presence of divalent ions, as the attraction between equally
charged plates \cite{bohinc}.}, but the simpler case is more
instructive. First of all, we note that the expression
Eq.~\eqref{last} represents only the pressure due to the inhomogeneous
electrolyte solution between the plates and we have therefore to add
the pressure contribution that comes from the homogeneous electrolyte
solution surrounding the system. Note that, since it has been assumed
that our system is electrically neutral, this osmotic contribution is
homogeneous in the surrounding space. Let $n_{b,1}=n_{b,-1}=n_b$ be
the bulk concentrations for the positive and negative monovalent ions.
We thus introduce the {\em excess} pressure $P$ that satisfies
\begin{eqnarray}
 P &=& \Pi(x) - \Pi_{\infty} = \nonumber \\ 
&=& -\frac{k_B T}{8 \pi \ell_B}\left( \frac{d\overline{\psi}}{dx} \right)^2 
+ k_B T \sum_{\alpha} (\overline{n}_{\alpha}(x)-n_b) \, 
\label{pression2}
\end{eqnarray}
and vanishes when $L$ increases toward infinity. The reservoir ions
are here assumed to behave like an ideal gas, coherently with the mean
field approximation.

Now, $P$ is a function of $L$, that can be obtained by solving the PB
Equation~\eqref{eqn20} for each $L\:\in\:\:]0;+\infty[$.

We then numerically analysed the sign of $P$ as a function of $L$ and
of the plate charge ratios $r=\sigma_L/\sigma_0$.  For $r>0$ (charges
of same sign) the plates always repel each other, which is a general
consequence of the PB theory \cite{Parsegian}.  For the particular
case $r=-1$ the interaction is instead always attractive.
Interestingly, in the more general case of $r<0$, and $r\neq -1$,
there always exists one and only one {\em equilibrium distance}
$L_{min}(r)$ between the plates for which we observe a transition
between attraction and repulsion (i.e. a vanishing $P$).  The
transition occurs at a distance that depends on the charge densities
of the plates, and a pronounced repulsion always appears at short
distances, despite the fact that the plates are oppositely charged.

Such transitions were already predicted in linearized treatments of
the problem \cite{Parsegian} in 1972. More recently, the non linear
case has been reconsidered \cite{Ben-Yaakov}, although an exact
derivation of the transition distance as a function of the plates
charges and salt concentration had already been obtained by Ohshima
\cite{Ohs75} in 1975.

Following the main lines of Ref.~\cite{Ohs75}, it is possible to
obtain an analytic expression for the equilibrium position explicitly
dependent on the plate charge densities and on the salt concentration,
for the case of a monovalent solution. We perform this calculation
explicitly in Appendix A. We obtain the following expression for the
position of the energy minimum $L_{min}$:
\begin{equation}
L_{min} = \lambda_D 
\Big\vert 
\ln{\Big( 
\frac{|\sigma'_0| (2+\sqrt{{\sigma'_L}^2+4})}{|\sigma'_L| (2+\sqrt{{\sigma'_0}^2+4})}
\Big) } 
\Big\vert
\label{Lmin}
\end{equation}
where we have introduced the Debye length, $\lambda_D=1/\sqrt{8 \pi
\ell_B \;n_b}$ and the adimensional  charge densities
${\sigma'}_0 =   4 \pi \ell_B  \lambda_D \sigma_0 $ and
${\sigma'}_L \; =\; \;\;  4 \pi \ell_B  \lambda_D \sigma_L$. A 
similar expression for the distance at which $P=0$ is given 
in Ref.~\cite{Ben-Yaakov}, Eq.~(9)
\footnote{
We noticed a difference of sign in the definition 
of the parameters $\gamma_\pm$ of Ref.~\cite{Ben-Yaakov}
with respect to our notation. This discrepancy arises from 
a different choice of the boundaries at which the pressure is 
evaluated, and has no consequences on the results, provided 
that the absolute value of the equilibrium distance $d$ 
($=L_{min}$) is taken in equation (9) of Ref.~\cite{Ben-Yaakov}.
for the case where $\sigma_0>\sigma_L$.
}.

\begin{figure}[h!]
  \begin{center} 
\includegraphics[width=0.4\textwidth]{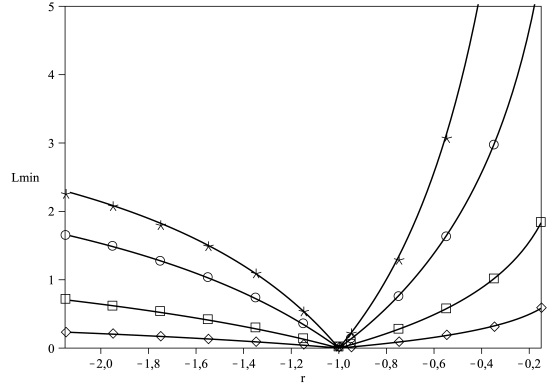}
\caption{Comparison between the estimation for the
position of the energy minimum of Eq.~\eqref{Lmin} (solid
lines) and the values of $L_{min}$ obtained on the basis of the direct
resolution of the Poisson-Boltzmann model (points). We chose 
$\sigma_0=-0.05\, e/$nm$^2$.  Lengths are given in nm, and as
functions of the ratio $r=\sigma_L/\sigma_0$, for different salt
concentrations: $0.001$ M (asterisks), 0.01 M (circles), 0.1 M
(squares), 1 M (diamonds).}
\label{fig:formula}
  \end{center}
\end{figure}

In Fig.~\ref{fig:formula} we compare the previous expression
Eq.~\eqref{Lmin} for $L_{min}$ with the corresponding values directly
obtained from the numerical solution of the PB equation, for different
salt concentrations and charge density ratios.
The minimum positions $L_{min}$ are numerically estimated
directly from the energy profiles. As expected, the
formula of Eq.~\eqref{Lmin} exactly agrees with the numerical results.

Two    limiting   regimes  can   now  be   considered.  
Let introduce the
Gouy-Chapman lengths  for both plates,  $\lambda_0 =  |1/ 2 \pi \ell_B
\sigma_0|$ and $\lambda_L = |1/ 2 \pi
\ell_B \sigma_L|$.
In low salt conditions, $\lambda_D \gg \lambda_0$ and $\lambda_D \gg
\lambda_L$. As a consequence, the position of the energy minimum is
approximately given by
\begin{equation}
L_{min} \simeq 2 \lambda_D \Big\vert \frac{1}{|{\sigma'}_L|} -
\frac{1}{|{\sigma'}_0|} \Big\vert = \vert \lambda_L - \lambda_0 \vert
\;\;\;\;{\rm (low \;\;\; salt),}
\end{equation}

In this limit, $L_{min}$ becomes therefore independent of the salt
concentration and is only a function of the plates charge densities,
i.e. of the ratio $r$ for the cases considered here since $\lambda_0$
is always kept fixed. In Fig.~\ref{fig:conc}, we report the local ion
distribution in the inter-plate space as a function of $x/L$ for a
given choice of the plate charges and for different bulk ion
concentrations $n_b$.  At low salt (Fig.~\ref{fig:conc} a), the
concentration of the counter-ions of the most charged plate is much
larger than the salt concentration.  In this case, the short range
repulsion is therefore mainly due to the counter-ions of the most
charged plate.  We stress that the solid and dotted curves in
Figure~\ref{fig:formula} also correspond to this low salt regime.

\begin{figure}[ht!]
  \begin{center} 
\includegraphics[width=0.2\textwidth]{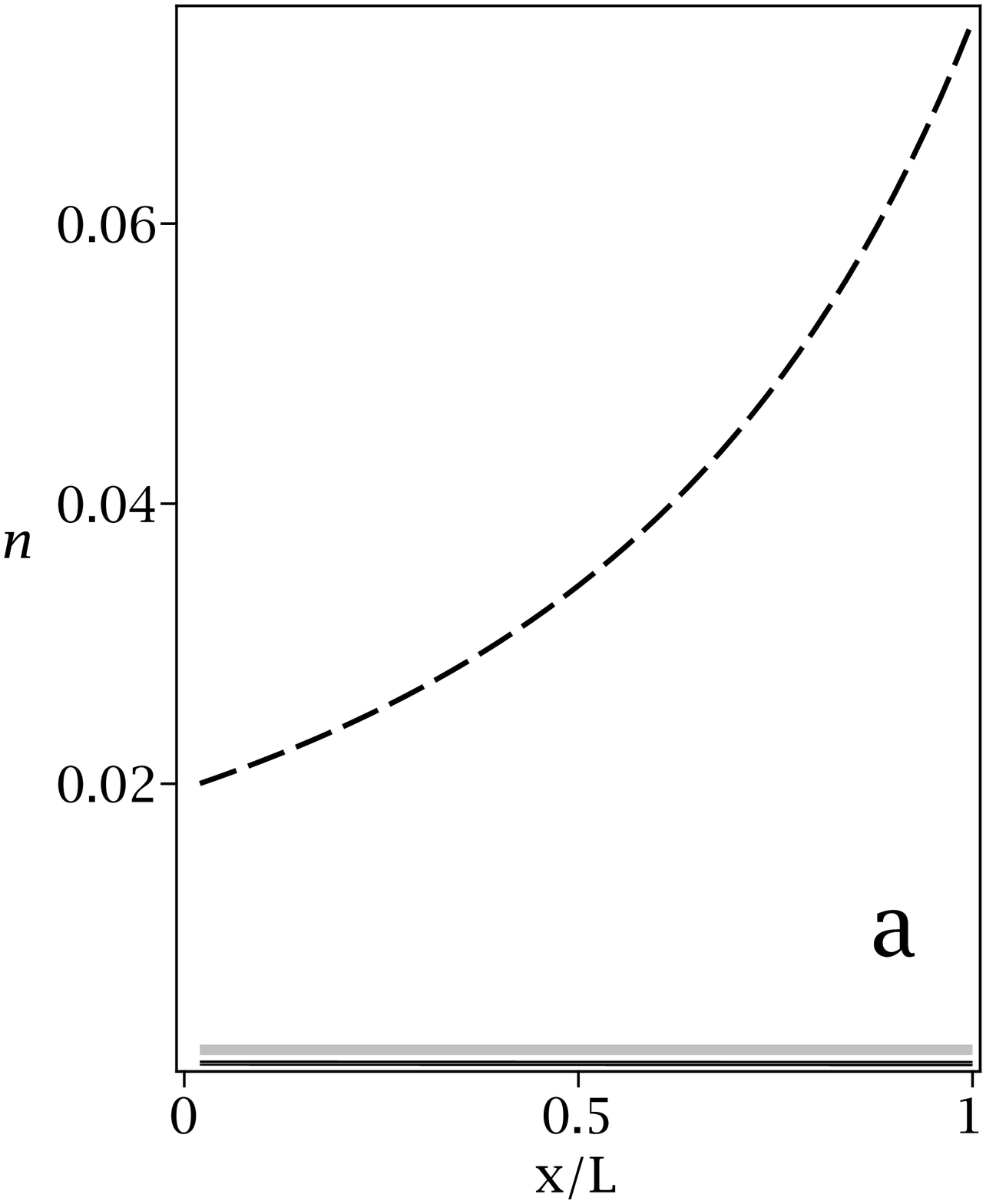}
  \includegraphics[width=0.2\textwidth]{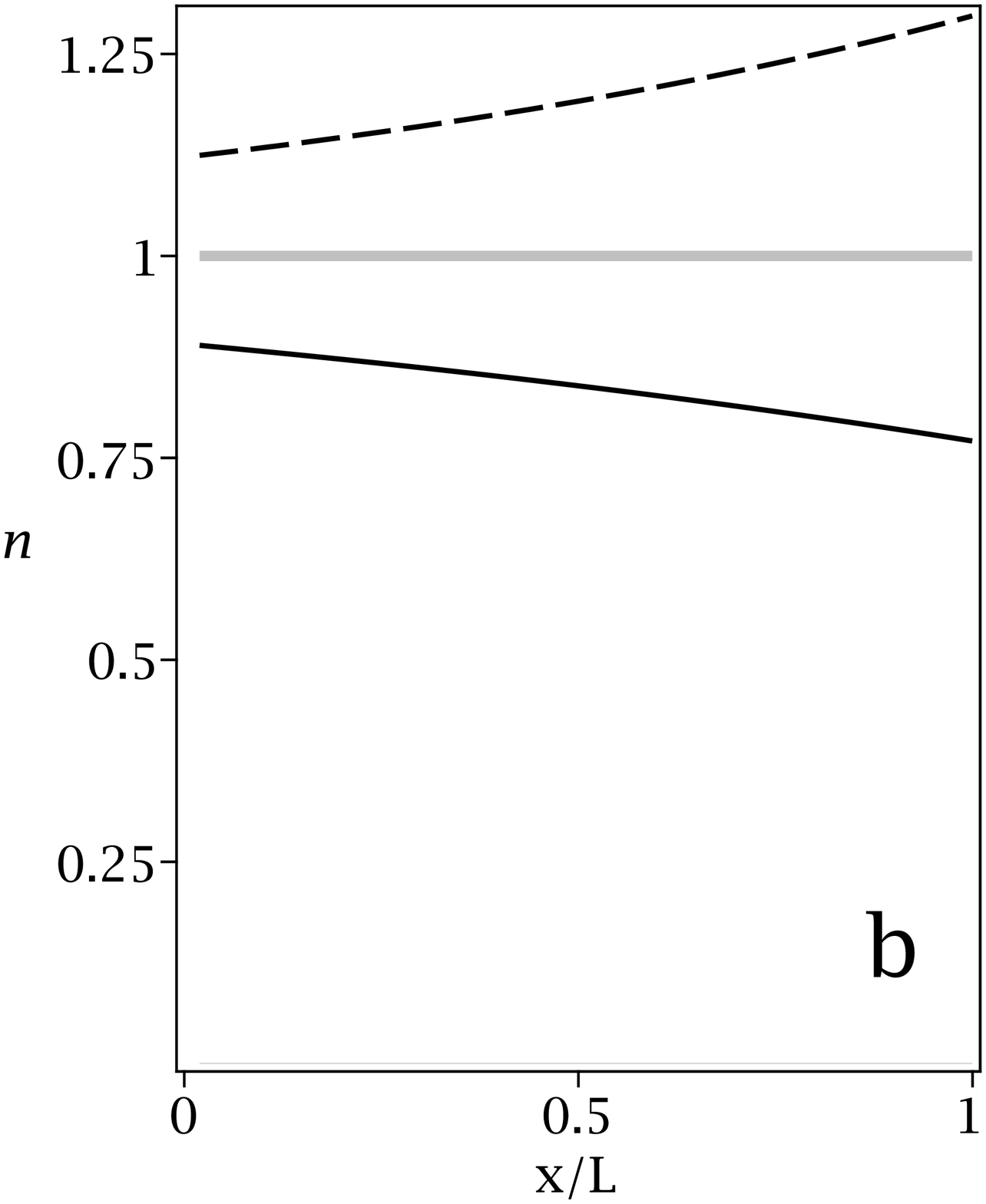} 
\caption{Positive (solid line) and negative (dashed line)
ion distributions in the inter-plate space, at the equilibrium
distance $L=L_{min}$, for the case of two charged plates with charge
densities $\sigma_0=-0.05\, e/$nm$^2$ and $\sigma_L=0.1\, e/$nm$^2$,
and for two different salt concentrations $n_b$: 0.001 M (low salt,
{\bf a}) and 1 M (high salt, {\bf b}). The thick grey line indicates
the value of the bulk ion concentration $n_b$.}
\label{fig:conc}
  \end{center}
\end{figure}

Inversely, at high salt, $\lambda_D \ll \lambda_0$ and $\lambda_D \ll
\lambda_L$, and the equilibrium is
\begin{equation}
L_{min} \simeq
 \lambda_D \;\; \Big\vert  \ln{\vert \frac{{\sigma'}_0}{{\sigma'}_L} \vert} \Big\vert 
= \lambda_D \;\; \Big\vert  \ln{\vert \frac{{\sigma}_0}{{\sigma}_L} \vert} \Big\vert 
\;\;\;\;{\rm (high \;\;\; salt).}
\end{equation}
In this limit the estimated equilibrium length $L_{min}$ is then
proportional to the Debye length, i.e. to $n_b^{-1/2}$. As shown in
Fig.~\ref{fig:conc} b, the short range repulsion is indeed
essentially due to the salt ions whose osmotic effect is modulated by
the charges on the plates. The dashed and dot-dashed curves in
Figure~\ref{fig:formula} correspond to the high salt regime.  In this
high salt regime, a good approximated expression for the equilibrium
position can also be obtained in the framework of the linearised PB
equation, as expected. We checked indeed that the resulting expression
\cite{Parsegian} matches well the curves in Fig.~\ref{fig:formula}
for any ratios $r$ at high salt concentrations (data not
shown). Instead, the linear PB approximation cannot reproduce the
observed behavior at low salt, whereas the expression
of Eq.~\eqref{Lmin} remains exact.

\begin{figure}[ht!]
  \begin{center}
  \includegraphics[width=0.4\textwidth]{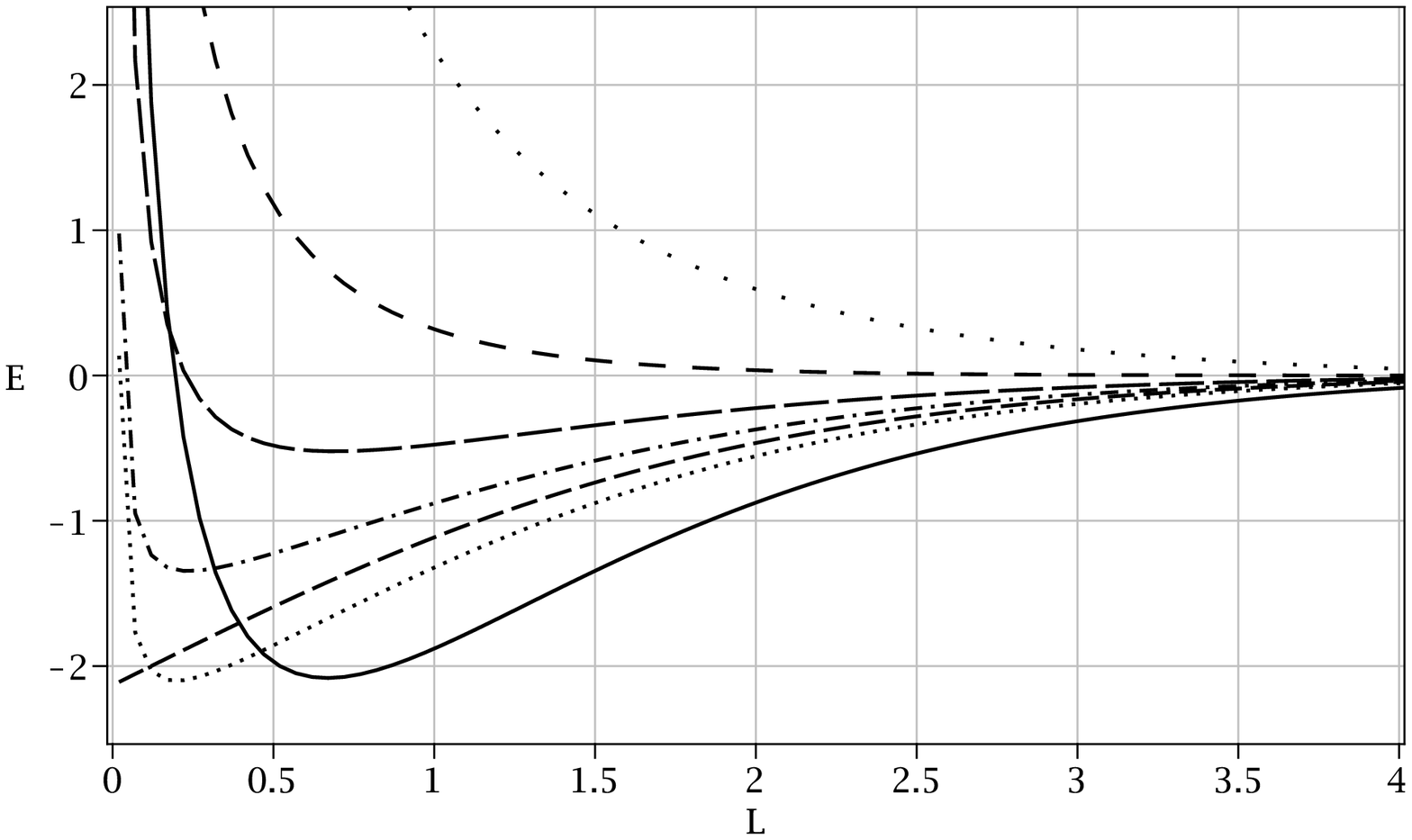} \\
  \fbox{\includegraphics[width=0.4\textwidth]{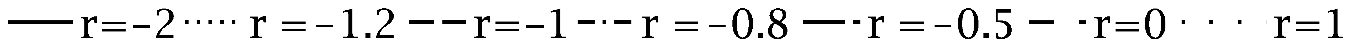}} 
\caption{The interaction energy per unit
surface $E$ ($k_B T/$100 nm$^2$) for the interaction of a plate of
charge density $\sigma_0=-0.05\, e/$nm$^2$ at $x=0$ with different
plates of charge densities $\sigma_L$, as a function of the distance
$L$ (nm) between them. The ratio $r=\sigma_L/\sigma_0$ between the
densities varies from $-2$ to $1$ according to the figure legend. The
plates are immersed in a 0.1 M monovalent solution.}
\label{fig:pressionenergie}
  \end{center}
\end{figure}

How the $P=0$ condition should be interpreted in terms of
electrostatic and osmotic contributions?  The mechanism leading to an
equilibrium position is not difficult to understand, starting from the
expression of the excess pressure, Eq.~\eqref{pression2}. As already
observed, the pressure is proportional to the constant of motion $C$
and is therefore constant in the inter-plate space $x \in [0,L]$.  Let
then consider the pressure $P$ exerted {\em on one plate}, e.g. at
$x=0$. In this limit, the electrostatic term in Eq.~\eqref{pression2}
simply reads $- {k_B T}\; \sigma_0^2\; /2$, and is therefore
independent of $L$. On the contrary, the osmotic term depends on the
ion concentration and is therefore a function of $L$. We then note
that the expression for the ion density, Eq.~\eqref{eqn16} can be
rewritten in terms of the bulk concentrations and of the reduced
potential as
\begin{equation}
\overline{n}_{\alpha}(x)= {n}_b e^{- z_{\alpha}\overline{\psi}(x)} \,. 
\label{ion}
\end{equation}
The equilibrium condition $P=0$, when calculated both on the left and
right plates, can thus be written as a condition for the mean field
potential at the boundaries that reads
\begin{eqnarray}
&&\cosh{\overline{\psi}_0(L_{eq})} 
=  \frac {  \pi \ell_B   \sigma_0^2}  {n_b}  + 1  
\,.
\label{equilibrium1} \\
&&\cosh{\overline{\psi}_L(L_{eq})} 
=  \frac {  \pi \ell_B   \sigma_L^2}  {n_b}  + 1  
\,.
\label{equilibrium2}
\end{eqnarray}
where we introduced $\overline{\psi}_0(L)=\overline{\psi}(x=0)$ and
$\overline{\psi}_L(L)=\overline{\psi}(x=L) \;\;
\forall\:L\:\in\:\:[0,\infty[$, and where $L_{eq}$ is a distance
between the plates for which $P=0$.

\begin{figure}[ht!]
  \begin{center} 
  \includegraphics[width=0.4\textwidth]{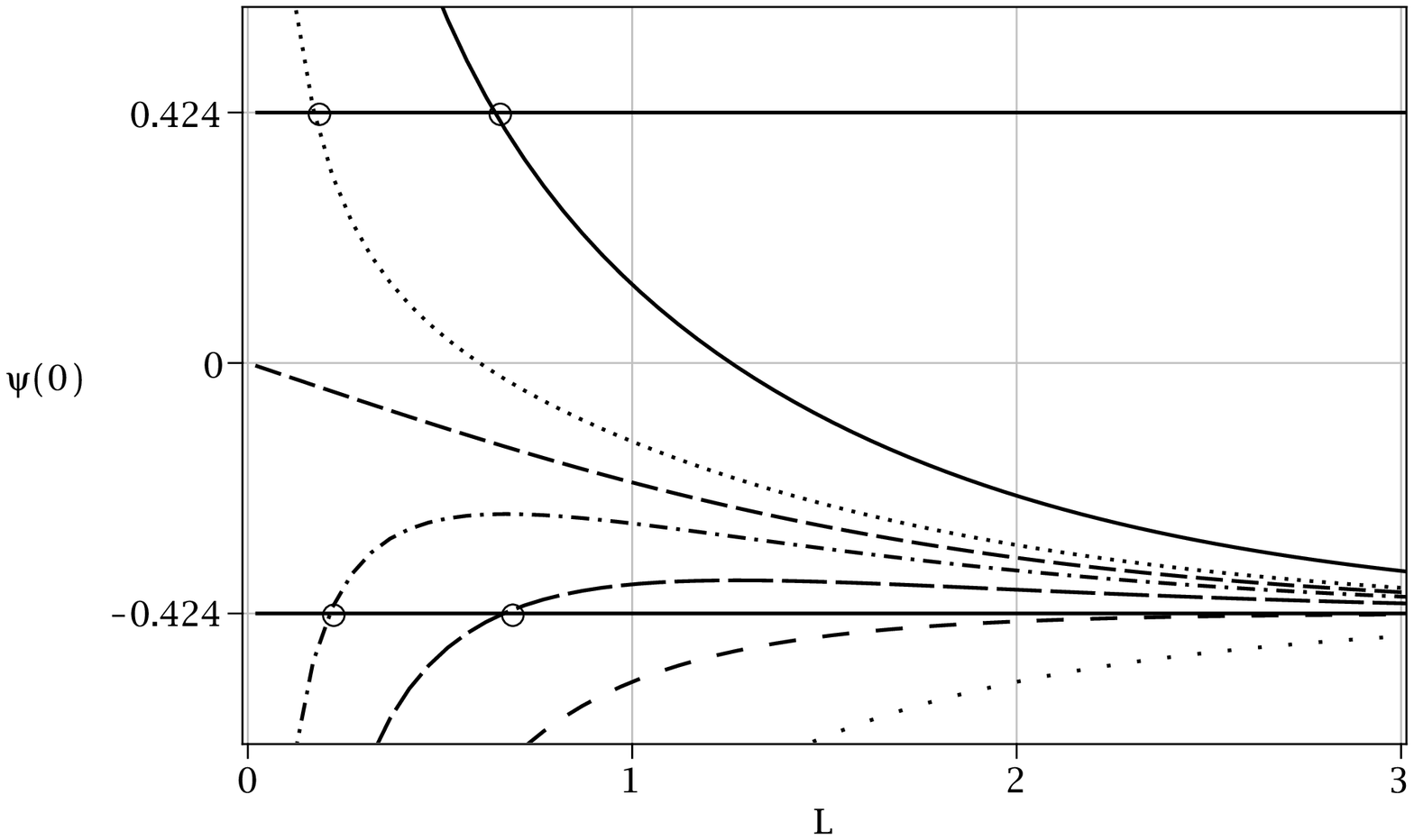}\\
  \fbox{\includegraphics[width=0.4\textwidth]{LEGEND_L.eps}}
\caption{The adimensional mean field potential at $x=0$,
$\overline{\psi}(0)$, as a function of the inter-plate distance $L$
(nm) and for the same ionic strength and charge density ratios of
Fig.~\ref{fig:pressionenergie}. The two horizontal lines correspond to
the two roots $\pm$ arccosh${\left( \pi \ell_B \sigma_0^2/ {n}_b
\,+\,1 \right)}$ of the left plate equilibrium condition,
Eq.~\eqref{equilibrium1}. The circles emphasize the non trivial
solutions leading to an equilibrium position.}
\label{fig:psi0}
  \end{center}
\end{figure}

From now on we will focus on the case of asymmetrically and oppositely
charged plates, $r<0$. In Figure~\ref{fig:psi0} we compare the
adimensional potential $\overline{\psi}_0(L)$ to the two roots
$\pm$arccosh${\left( \pi \ell_B \sigma_0^2/ {n}_b \,+\,1 \right)}$ of
Eq.~\eqref{equilibrium1}. A trivial solution to
Eqs. \eqref{equilibrium1} and \eqref{equilibrium2} corresponds to the
limit $L \rightarrow \infty$, where $P=0$ by construction.  In this
limit, and typically for $L \gg \lambda_0+ \lambda_L$, each plate
charge is neutralized by its cloud of counterions as if the other
plate didn't exist. At a large enough distance $x$ from the plates,
therefore, the ionic atmosphere behaves as an ideal gas and its
pressure contribution $\Pi(x)$ exactly compensates the reservoir
pressure $\Pi_\infty$.  We also note that, in this case, according to
\eqref{ion}, the sign of $\overline{\psi}_0$ and $\overline{\psi}_L$
at infinity are opposite for oppositely charged plates.

Let now look for another, finite solution of Eqs. \eqref{equilibrium1}
and \eqref{equilibrium2} leading to $L_{eq}=L_{min}$. For this case,
we only have to determine the sign of $\overline{\psi}_0(L_{min})$ and
$\overline{\psi}_L(L_{min})$.  It easy to see that the only possible
solution is that the counterions of one plate prevail on {\em both}
plates, so that the signs are both equal.  It is the potential of the
less charged plate that will change its sign at a given distance $L$
between infinity and $L_{min}$.  In Figure ~\ref{fig:psi0} we show
indeed that the sign of the adimensional potential $\overline{\psi}_0$
changes for $r<-1$ whereas it remains the same when $r>-1$. The
opposite arises for $\overline{\psi}_L$ (data not shown).

The physical interpretation of the observed equilibrium is thus
straightforward. The most charged plate carries its cloud of condensed
counterions when approaching the less charged one. The counterions of
the less charged plate are more easily released in the bulk when the
two ion clouds overlap. This process continues until the
electroneutrality constraint precludes a further release of ions, this
leading the ion concentration, and therefore the repulsive osmotic
pressure, to increase. The equilibrium is then obtained at the
distance $L_{min}$ where electrostatic attraction and osmotic
repulsion exactly balance.

\begin{figure}[ht!]
  \begin{center} 
\includegraphics[width=0.4\textwidth]{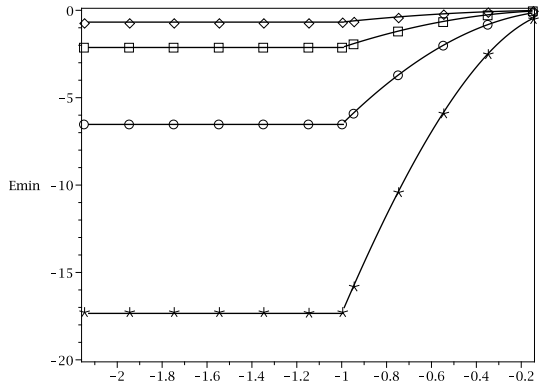}
\caption{Comparison between the estimation 
of the energy minimum of Eq.~\eqref{Emin} (solid
lines) and the values of  the energy at the minimum position
$L_{min}$ obtained by direct integration of the Poisson-Boltzmann
model, for different salt concentrations: $0.001$ M (asterisks), 0.01
M (circles), 0.1 M (squares), 1 M (diamonds). We have udes again
$\sigma_0=-0.05\, e/$nm$^2$.  In order to use more natural units,
energies are given in units of $k_B T/$ 100 nm$^2$.}
\label{fig:energies}
  \end{center}
\end{figure}

\subsection{Energy value $E_{min}$ at the minimum}

We have shown that the pressure always vanishes at a given inter-plate
distance for oppositely, asymmetric charged plates. Nevertheless, if
the presence of a vanishing pressure always corresponds in principle
to an equilibrium position between the plates, it does not guarantee
by itself that this equilibrium position will be stable enough to be
relevant from a thermodynamic point of view.  Indeed, in order to
assess the real existence of a stable equilibrium, we need to estimate
the corresponding energy gain. We stress again that, if the energy
gain at the minimum is small with respect to $k_B T$, the two charged
bodies will behave as in the absence of electrostatic interaction (the
energy going to zero at large distances). On the contrary, a deep
minimum will make the bodies stabilize at a non-zero equilibrium
distance.

The explicit calculation of the whole
function $P(L)$ allows us to evaluate the energy profile $E(L)$ and
compare the depth of the potential well to the thermal energy $k_B
T$. The energies per unit area are shown in
Fig.~\ref{fig:pressionenergie} (bottom), for the same charge densities
as considered above. In order to use more natural units, energies are
given in units of $k_B T/$ 100 nm$^2$.  As expected, an energy minimum
always exists for $r<0$ and $r\neq -1$.  Interestingly, while for
$0<r<1$ the energy minimum depth shows a relevant dependence on the
second plate charge density $\sigma_L$, this dependence disappears for
$r<-1$ where the depth becomes constant.

A more systematic investigation of the energy minimum depth for
varying charge densities and salt conditions is shown in
Figure~\ref{fig:energies}. The value of the energy per a unit area of
100 nm$^2$ for different ionic strengths is given as a function of
the ratio of the plates charge densities $r$, again for
$\sigma_0$ fixed at $-0.05\, e/$nm$^2$. In low salt, the energy
minimum depends on the ratio $r$ and on the salt concentration, and it
reaches its maximum value for $r=-1$, i.e. when the position of the
minimum degenerates to $L=0$. Energy depths are up to roughly $10$
$k_B T$ per 100 nm$^2$. Figure~\ref{fig:energies} also confirms that
the minimum depth becomes constant for $r < -1$, and coincides in this
case with its (maximum) value at the singular value $r=-1$.

In his paper, Ohshima \cite{Ohs75} also obtains an analytic expression
for the energy at the minimum. An equivalent calculation, adapted to our
formalism, is presented in Appendix B. The final result reads
\begin{equation}
\beta E_{min}=
 8 n_b \lambda_D \;
\left\lbrace \sqrt{ |{\sigma'}_m|^2+4}-2
-|{\sigma'}_m| \; \mathrm{arcsinh}  \vert \frac{{\sigma'}_m}{2} \vert \right\rbrace \,,
 \label{Emin}
\end{equation}
where ${\sigma'}_m$ is the adimensional charge parameter related to
the {\em smallest} plate charge density, i.e. ${\sigma'_m} =
\min{({\sigma'_0}, {\sigma'_L})}$.  In Figure~\ref{fig:energies} we
compare the previous expression for the value of the energy depth with
the results obtained by direct integration of the Poisson-Boltzmann
equation. The two results show a perfect agreement. Together with
Equation~\eqref{Lmin}, the last result allows to a rapid and precise
estimate of the equilibrium position and strength, and represent
therefore a powerful tool in order to study the effective interaction
between charged bodies in solution.

\begin{figure}[ht!]
  \begin{center} 
\includegraphics[width=0.22\textwidth]{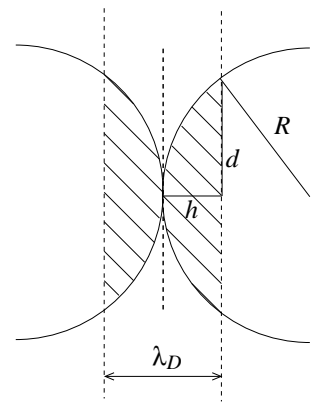}
\caption{Sketch for the calculation of the interacting
area for the case of two identical spheres. The area of each dashed
spherical surface is $S= \frac{\pi}{4}(d^2 + 4 h^2)$.}
\label{fig:calotte}
  \end{center}
\end{figure}

Nevertheless, we have to stress finally that Equation~\eqref{Emin}, as
well as Figures~\ref{fig:pressionenergie} and~\ref{fig:energies}, only
give the energy {\em per unit area} (fixed to 100 nm$^2$ in the
Figures).  To obtain the total energy between two charged bodies in
solution and compare it to the thermal energy, the surface and
geometry of the bodies should be taken into account. This roughly
amounts to multiply the energy by an {\em effective interaction area},
but the estimation of this area is not easy in that it depends on the
bodies shape. Indeed, the variation of the interaction with the
distance should be included in the calculation of the effective
interaction area. A typical choice is the use of the Derjaguin
approximation \cite{Der34,Whi83}, that calculates the interaction
energy $U$ between two curved surfaces by integrating the interaction
energy per unit area between two flat plates $E(L)$ as
\begin{equation}
U \simeq \int_A E(L) dA \simeq f([R_1],[R_2]) \int_{D_{min}}^\infty E(L) dL \,,
\end{equation}
where $D_{min}$ is the distance of closest approach between the two
curved surfaces, $dA$ is the differential area of the surfaces facing
each other, $[R_1]$ and $[R_2]$ represent the sets of principal radii of
curvature of the surfaces 1 and 2, respectively, at the distance of
closest approach, and $f([a1],[a2])$ is a function of the radii of
curvature of the surfaces. 
A very rough estimate should consider that the interaction between the
two surfaces becomes negligible when their distance becomes larger
than the screening Debye length $\lambda_D$.   In a $0.1$ M solution, the Debye
length is of the order of 1 nm. For the case of two
spherical colloids of 100 nm diameter, (it varies from
approximately 0.3 nm for an 1 M solution to 10 nm for 0.01 M).
If the two spheres are in contact (i.e. for $L=0$), a
simple geometric construction (depicted schematically in
Fig.~\ref{fig:calotte}) allows to calculate the surface area on each
sphere that is separate by less than a Debye length from the facing
one, this leading to a surface of the order of 300 nm$^2$ (roughly 1
\% of the whole sphere surface). As a consequence, the depth of the energy 
well would reach approximately $\sim$20 $k_B T$ in these conditions,
and be thus large enough to ensure its stability at ordinary thermal
conditions. Note however, that the two spheres will {\em not} be in
contact anymore at equilibrium, but at a distance close to $L_{min}$,
which can vary from $0$ to approximately $2$ nm depending on the value
of the ratio $r$.  When this distance becomes comparable to
$\lambda_D$, the interaction area decreases considerably. Therefore,
the actual interaction surface will be in general reduced to a value
that depends on the two lengths $L_{min}$ and $\lambda_D$, and should
be calculated case by case.  Note that, in this sense, the shape of
the two interacting bodies is bound to play a crucial role, in that it
can modify considerably the distances between the facing surfaces and
consequently the effective interaction area.

\section{Summary and conclusion}

In this paper we derived, within the genuine framework of the
Poisson-Boltzmann approximation, the main quantities of interest for
the interaction between two charged surfaces in solution.

Interestingly, such a rigorous derivation brings up an extended
formula for the interplate pressure, formally differing from the usual
expression (giving the sum of electrostatic and osmotic contributions)
by an extra-term.  While in the considered case of fixed charges
boundary conditions, this additional term vanishes and we recover the
standard expression for the pressure, the question arises whether it leads
to a modified result when different boundary conditions are
considered.  We recalled  that the Poisson-Boltzmann free
energy is properly used as the reference thermodynamic potential
{\em only} at fixed charges. To show how the choice of the thermodynamic
potential depends on the boundary conditions, we explicitely solved
the problem in the fixed potential case, and show  how to recover 
the physical meaningful pressure in that case. These precisions on the
interplay between boundary conditions and thermodynamic ensemble
have been obtained here thanks to a detailed and deductive derivation
from the very basis of the Poisson-Boltzmann approach.  Let us stress once
again that shortcutting the details of the mathematical derivation
potentially leads to underappreciate the role of the boundary
conditions.

We then explicitely solved the problem in the simple case of a 1:1
salt solution and observed a very rich behavior as a function of the
ratio of the plates charge densities and of the salt concentration,
with a non trivial equilibrium position arising in large intervals of
these parameters. The distance at which the osmotic and electrostatic
pressures are in equilibrium is finely tuned by the system
parameters. Such equilibrium position can stabilize two asymmetrically
charged bodies at a nonzero distance, provided that the corresponding
free energy gain is large enough compared to the thermal noise.  At a
given temperature, the stability of the complex can therefore be
assessed only by explicitly calculating the depth of the corresponding
energy well. We obtained a readily available answer to this problem by
deriving analytic expressions for the position and depth of the energy
well, by a reactualized version of the overlooked derivation of
Ohshima \cite{Ohs75}.  

In order to compare the energy values with the thermal energy, an
estimation of the interaction area is also necessary. As an example,
we gave an estimation for the typical case of spherical colloids, and
found that the interaction energy well is typically of the order of
several $k_B T$, this leading to a quite deep minimum. This estimation
is only a rough approximation because, for a given problem either in
biological or colloidal systems, the behavior of the two interacting
bodies is strongly dependent not only on salt conditions and on the
bodies charge but also on their shape.    The relevance of the
 minimum  is consequently rather delicate to compute  and
model-dependent. \\


\renewcommand{\thesection}{\Alph{section}}
\setcounter{section}{0}
\setcounter{equation}{0}
\numberwithin{equation}{section}
\numberwithin{figure}{section}
\renewcommand{\theequation}{\thesection\arabic{equation}}
\renewcommand{\thefigure}{\arabic{figure}}


{\centerline{\large{\bf APPENDICES}}}

\vspace{-5mm}

\section{Determination of the minimum position.}

Starting from Eq.~\eqref{eqncalF2} and taking into account the
contribution of the surrounding ions as in Eq.~\eqref{pression2}, one
can obtain a suitable form for the {\em excess free energy} that will
allow one to evaluate its amplitude at a given position $x$.
We start by writing the excess free energy as
\begin{eqnarray}
 \beta \mathcal{F}_{PB}[\sigma,\{\overline{n}\}]
&=& 
\int_0^L \left[ \frac{1}{8\pi \ell_B} \left( \frac{ d \overline{\psi}}{dx} \right)^2 
- \sum_{\alpha=1}^2 \overline{n}_{\alpha}(x) \right] \;dx 
\nonumber \\&& +\, 2 n_b \, L 
\,+ \, \sum_{\alpha=1}^2 \int_0^L \overline{n}_{\alpha}[-z_{\alpha} \overline{\psi}(x)]\; dx 
\end{eqnarray}
where the $2 n_b \, L$ term arises from the $\Pi_\infty$ contribution.
Now, using \eqref{eqnE} and \eqref{eqn20} one finds:
\begin{eqnarray}
\beta \mathcal{F}_{PB}[\sigma \{ \overline{n} \}]
&=& 
2n_b \left\lbrace \left( \frac{1}{2}C'+1\right)L
+ \int_0^L \overline{\psi}(x) \sinh \overline{\psi}(x) \; dx \right\rbrace 
\nonumber \\ 
&=& 2n_b \left\lbrace \left( \frac{1}{2}C'+1\right)L
+ \lambda_D^2 \left[ \overline{\psi}(x)  \frac{d \overline{\psi}}{dx} \right]_0^L \right. 
\nonumber \\
&& \hspace{2cm}- \left. \lambda_D^2 \int_0^L \left( \frac{d \overline{\psi}}{dx} \right)^2 dx 
\right\rbrace 
\label{Ia}
\end{eqnarray}
where we have introduced the Debye length, $\lambda_D=1/\sqrt{8 \pi
\ell_B \;n_b}$, $C'=C/(4 \pi \ell_B n_b)$ and we performed an integration by part for the last equation.

In order to calculate $L_{min}$, it is convenient to introduce the
following adimensional parameters:
\begin{eqnarray}
&& \eta =  x /\lambda_D 
\label{eta} \\ 
&& \gamma(\eta)=\overline{\psi}(x) 
\label{gamma} \\ 
&& \theta(\eta)=\frac{d \gamma}{d \eta} \,.
\label{theta}
\end{eqnarray}
 In terms of these variables, the Poisson-Boltzmann
equation just writes
\begin{eqnarray}
&& \frac{d \theta}{d \eta} = \sinh \gamma \label{pbnew} 
\end{eqnarray}
with the boundary conditions
\begin{eqnarray}
&& \theta(0)=-{\sigma'}_0 =           - 4 \pi \ell_B  \lambda_D \sigma_0 \nonumber \\
&& \theta(\eta_L)={\sigma'}_L \; =\; \;\;  4 \pi \ell_B  \lambda_D \sigma_L
\label{bcnew} 
\end{eqnarray}
and the equivalent of Eq.~\eqref{eqnE}, giving the constant of motion
for the system, reads
$\theta^2 = 2 \cosh \gamma + C'$.
From this equation one gets
\begin{eqnarray}
&& \frac{( \theta^2-C')^2}{4}= \cosh^2 \gamma= 1+ \sinh^2 \gamma
\nonumber \\ && \frac{( \theta^2-C')^2}{4}-1= \left(
\frac{ d \theta}{d \eta} \right)^2 \nonumber \\ & \Rightarrow &
\left| \frac{ d\eta }{d \theta} \right| =
\frac{1}{\sqrt{(\theta^2-C')^2/4-1}}
\end{eqnarray}
By integrating $\eta$ from $0$ to $\eta_L$, with $\left|d \eta \right|
= d \eta$, we thus obtain 
\begin{eqnarray}
\int_{\theta(0)}^{\theta(\eta_L)} \left| \frac{ d\eta }{d \theta} \right|\; d\theta 
= \int_0^{\eta_L} d\eta \frac{d\theta}{|d\theta|} \,. \nonumber
\end{eqnarray}
We are interested here in the case of oppositely charged plates, where
the energy minimum does exist. We have therefore to consider different
cases. Let assume for the moment that $|{\sigma'}_0|>
|{\sigma'}_L|$. If now ${\sigma'}_0 < 0$ and ${\sigma'}_L > 0$, then
$\theta(0)$ and $\theta(\eta_L)$ have both positive values and
$|d\theta | = - d\theta $. We then have
\begin{eqnarray}
\eta_L = \int_{|{\sigma'}_L|}^{|{\sigma'}_0|} 
\left| \frac{ d\eta }{d \theta} \right| \; d\theta 
\hspace{1cm}  {\sigma'}_0 < 0,\;\;{\sigma'}_L > 0
\,.
\label{Lmin0}
\end{eqnarray}
On the other hand, if ${\sigma'}_0 > 0$ and ${\sigma'}_L < 0$, then
$\theta(0)$ and $\theta(\eta_L)$ have both negative values and
$|d\theta | = d\theta $.
We thus obtain 
\begin{eqnarray}
\eta_L = -\int_{-|{\sigma'}_L|}^{-|{\sigma'}_0|} \left| \frac{ d\eta }{d \theta} \right| \; d\theta 
\hspace{1cm}  {\sigma'}_0 > 0,\;\;{\sigma'}_L < 0
\,.
\label{Lmin1}
\end{eqnarray}
Nevertheless, since $\left| \ d\eta/ d \theta \right|$ is only a
function of $\theta^2$ we can make the change of variable $\Theta =
-\theta$ in Eq.\eqref{Lmin1} and  retrieve the result \eqref{Lmin0}.
Therefore, 
 we always have 
\begin{eqnarray}
\eta_L = \int_{|{\sigma'}_L|}^{|{\sigma'}_0|} \frac{ 2 }{\sqrt{(\theta^2-C'+2)(\theta^2-C'-2)}}\; d\theta 
\label{Lmin2}
\end{eqnarray}

If now $L={L_{min}}$, then $P=0$, i.e. $\Pi=\Pi_\infty=2 k_B T n_b$,
and therefore $C'=-2$. At the equilibrium position, Eq.\eqref{Lmin2}
becomes then
\begin{eqnarray}
\eta_{L_{min}}=\int_{|{\sigma'}_L|}^{|{\sigma'}_0|} \frac{ 2 }{ \theta \sqrt{ \theta^2+4}} \; d\theta 
= \int_{|{\sigma'}_L|/2}^{|{\sigma'}_0|/2} \frac{ 1}{ \alpha \sqrt{ \alpha^2+1}} d \theta \nonumber 
\end{eqnarray}
where we introduced $\alpha = \theta/2$. As a primitive function of
$1/(x\sqrt{x^2+1})$ is $\ln \left( x/(1+\sqrt{x^2+1}) \right)$, we get
for the case $|{\sigma'}_0|> |{\sigma'}_L|$
\begin{equation}
\eta_{L_{min}}=
\ln \left( \frac{ |{\sigma'}_0|(2+ \sqrt{|{\sigma'}_L|^2+4})}{|{\sigma'}_L|(2+ \sqrt{|{\sigma'}_0|^2+4})} 
\right) \,.
\end{equation}

The extension to the opposite case of $|{\sigma'}_0|<|{\sigma'}_L|$
is straightforward, this leading to the following final formula for the
position of the energy minimum $L_{min}=\lambda_D \eta_{L_{min}}$:
\begin{eqnarray}
L_{min} = \lambda_D 
\Big\vert 
\ln{\Big( 
\frac{|\sigma'_0| (2+\sqrt{{\sigma'_L}^2+4})}{|\sigma'_L| (2+\sqrt{{\sigma'_0}^2+4})}
\Big) } 
\Big\vert \,. \nonumber
\end{eqnarray}

\section{Determination of the energy value at the minimum.}

Following the main lines of the calculation presented in \cite{Ohs75},
we here look for an analytic expression for the energy at the minimum.
In our framework, the interaction energy computed numerically directly from 
the integration of the excess pressure writes formally (by definition of the 
integration):
\begin{equation}
\beta E(L)=\beta ( \mathcal{F}(L)-\mathcal{F}(\infty) ) 
\label{Vmin0}
\end{equation}
where $\mathcal{F}(L)$ is given by \eqref{Ia}. By using the dimensionless 
parameters defined in the previous section, we have
\begin{eqnarray}
\beta \mathcal{F} &=& \left\lbrace (K(L)+J(L)-I(L) \right\rbrace  2n_b \lambda_D \,,
\label{Vmin1} 
\end{eqnarray}
with
\begin{eqnarray}
K(L)&=&(\frac{1}{2}C'+1)\eta_L \,, \nonumber  \\ 
J(L)&=&\Big[ \gamma(\eta) \theta(\eta) \Big]_0^{\eta_L} \,, \nonumber \\
I(L)&=& \int_0^{\eta_L} \theta^2\;  d\eta \,. \nonumber 
\end{eqnarray}

We will now, first, calculate the three contributions to
$\mathcal{F}(L_{min})$, and then the corresponding contributions to
$\mathcal{F}(\infty)$.

Let start by calculating $I(L_{min})$. From the definition, we have
\begin{eqnarray}
I(L_{min})&=&\int_0^{\eta_{L_{min}}} \theta^2 \left(\frac{d \eta}{d \theta} d \theta \right) 
= -\int_{|{\sigma'}_0|}^{|{\sigma'}_L|} \theta^2 \left|\frac{d \eta}{d \theta} \right| d \theta \nonumber \\ 
&=& \int_{|{\sigma'}_L|}^{|{\sigma'}_0|} \frac{2 \theta }{ \theta^2+4} d\theta \nonumber \\ 
&=& 2( \sqrt{|{\sigma'}_0|^2+4}-\sqrt{|{\sigma'}_L|^2+4}) 
\label{Vmin4}
\end{eqnarray}
where we used the fact that, on the $[0,\eta_L]$ range, $d\eta/d\theta
= - |d\eta/d\theta|$.

To calculate $J(L_{min})$ we recall that when $L=L_{min}$ we have
$\cosh \gamma(0)= {{\sigma'}}_0^2/2 +1$. By using the relation $(\cosh x
-1)/2=\sinh^2({x}/{2})$, we then get $\sinh^2 \gamma(0)/2 
= {{\sigma'}}_0^2/4$, and thus
\begin{eqnarray}
 |\sinh \frac{\gamma(0)}{2} |= \frac{1}{2} | {{\sigma'}}_0 | 
\label{Vmin5} 
\end{eqnarray}
In the same way we have
\begin{equation}
|\sinh \frac{\gamma(\eta_{L_{min}})}{2} |= \frac{1}{2} | {\sigma'}_L | \,.
\label{Vmin6} 
\end{equation}
We can now easily calculate $J(L_{min})$ starting from
\begin{eqnarray}
J(L_{min})&=& \gamma(\eta_{L_{min}})\,\theta(\eta_{L_{min}})-\gamma(0)\, \theta(0) \nonumber \\
&=&\gamma(\eta_{L_{min}})\, |{\sigma'}_L|+\gamma(0)\,{\sigma'}_0 \,. \nonumber
\end{eqnarray}
Now, at $L_{min}$, $\gamma(\eta_{L_{min}})$ and $\gamma(0)$ have the
same sign, which is governed by the most charged plate. Let again
focus on the case where ${\sigma'}_0 < 0$ and
$|{\sigma'}_0|>|{\sigma'}_L|$, as used in our illustrations. In this
case, the sign of $\gamma(\eta_{L_{min}})$ and $\gamma(0)$ is the same
as the sign of ${\sigma'}_0$ (cf. Figure 4), this leading to
\begin{equation}
J(L_{min})
=- 2 \,|{\sigma'}_L| \; \mathrm{arcsinh} \vert \frac{{\sigma'}_L}{2} \vert
 - 2 \,|{\sigma'}_0| \; \mathrm{arcsinh} \vert \frac{{\sigma'}_0}{2} \vert
\label{Vmin7}
\end{equation}

We easily obtain that $K(L_{min})=0$ since  $C'=-2$ for $L=L_{min}$.

The overall result for $\mathcal{F}(L_{min})$ reads  therefore
\begin{eqnarray}
&&\beta \mathcal{F}(L_{min})=
4 n_b \lambda_D \;
\left\lbrace   
   - |{\sigma'}_L|\; \mathrm{arcsinh} \vert \frac{{\sigma'}_L}{2} \vert \right.
         \\  
&& - |{\sigma'}_0|\; \mathrm{arcsinh} \vert \frac{{\sigma'}_0}{2} \vert
 \left. -( \sqrt{|{\sigma'}_0|^2+4}-\sqrt{|{\sigma'}_L|^2+4} \,) \right\rbrace \,.
\nonumber
\hspace{8mm}
\label{Vmin8}
\end{eqnarray}

Let now calculate $ \mathcal{F}(\infty)$.  We have to be quite
cautious to compute $I(\infty)$. Indeed, we have
\begin{equation}
I(\infty)
= \lim_{\eta_L \rightarrow \infty} \int_0^{\eta_L} \theta^2 
\left( \frac{d \eta }{d \theta} d\theta \right) \,, \nonumber 
\end{equation}
and the point is that $d\eta/d\theta$ has not the same sign all over the
range $[0, \infty[$. Actually because of the infinite distance between
the two plates, each plate tends to behave as a single plate in this
limit, and thus there exists a distance for which $\theta =0$. So,
$\theta$ will be initially equal to $|{\sigma'}_0|$, then decrease to
zero and increase again to $|{\sigma'}_L|$. We have therefore:
\begin{eqnarray}
I(\infty)
&=&-\int_{|{\sigma'}_0|}^0 \theta^2 \left| \frac{d \eta}{d \theta} \right| \, d \theta 
+ \int_0^{|{\sigma'}_L|}\theta^2 \left| \frac{d \eta}{d \theta} \right| \, d \theta \nonumber \\ 
&=& -\left[ 2 \sqrt{\theta^2+4} \right]_{|{\sigma'}_0|}^0 
+ \left[ 2 \sqrt{\theta^2+4} \right]^{|{\sigma'}_L|}_0 \nonumber \\ 
&=& -8 + 2\sqrt{|{\sigma'}_0|^2+4}+2\sqrt{|{\sigma'}_L|^2+4} \,.
\label{Vmin9}
\end{eqnarray}

Besides, we have for $J(\infty)$:
\begin{eqnarray}
J(\infty)&=& 
\gamma(\infty)|{\sigma'}_L|- \gamma(0)| {\sigma'}_0| \\ 
&=& 2 \, |{\sigma'}_L| \; \mathrm{arcsinh} \vert \frac{{\sigma'}_L}{2} \vert    
   -2 \, |{\sigma'}_0| \; \mathrm{arcsinh} \vert \frac{{\sigma'}_0}{2} \vert    
  \nonumber
\label{Vmin10}
\end{eqnarray}
because it's only the reduced potential $\gamma$ corresponding to the
plate with the lowest charge (in absolute value) that changes its sign
between $L_{min}$ and $L_{\infty}$ (cf. again Figure 4).

When there is an infinite distance between the plates we also have
$C'=-2$ (intuitively because there is no more interaction between the
plates) and thus $K(\infty)=0$. 
We then have:
\begin{eqnarray}
&& \beta \mathcal{F}(\infty)= 
4 n_b \lambda_D \;
\left\lbrace 
      |{\sigma'}_L| \mathrm{arcsinh} \vert \frac{{\sigma'}_L}{2} \vert \right.
           \\ 
&& -  |{\sigma'}_0| \mathrm{arcsinh}  \vert \frac{{\sigma'}_0}{2} \vert
\left. -(-4 + \sqrt{|{\sigma'}_0|^2+4}+\sqrt{|{\sigma'}_L|^2+4}) \right\rbrace  
\,.
\nonumber
\label{Vmin11}
\end{eqnarray}

From the evaluation of Equation~\eqref{Vmin0} at $L=L_{min}$, we get
then finally the following expression for the energy at the minimum
in the case when  ${\sigma'}_0 < 0$ and $|{\sigma'}_0|>|{\sigma'}_L|$:
\begin{equation}
\beta E_{min}=  
8 n_b \lambda_D \;
 \left\lbrace \sqrt{ |{\sigma'}_L|^2+4}-2-
|{\sigma'}_L| \; \mathrm{arcsinh}  \vert \frac{{\sigma'}_L}{2} \vert
\right\rbrace \,.
\label{Vmin12}
\end{equation}
On the other hand, one can easily be convinced that inverting the
roles of $|{\sigma'}_0|$ and $|{\sigma'}_L|$ leads to the same
expression as Eq.~\eqref{Vmin12} where $|{\sigma'}_0|$ and
$|{\sigma'}_L|$ are inverted.  Therefore, the very general result
writes (Eq.~\eqref{Emin})
\begin{eqnarray}
&&\beta E_{min}=
 8 n_b \lambda_D \;
\left\lbrace \sqrt{ |{\sigma'}_m|^2+4}-2
-|{\sigma'}_m| \; \mathrm{arcsinh}  \vert \frac{{\sigma'}_m}{2} \vert \right\rbrace \,,\nonumber \\
\end{eqnarray}
where ${\sigma'}_m$ is related to the {\em smallest} plate charge
density, i.e. ${|\sigma'_m|} = \min{({|\sigma'_0|}, {|\sigma'_L|})}$.


\end{document}